\documentclass[english]{article}
\usepackage[T1]{fontenc}
\usepackage[utf8]{inputenc}
\usepackage{mathrsfs}
\usepackage{amsmath}
\usepackage{amssymb}
\usepackage{graphicx}
\usepackage{babel}
\begin{document}
\title{Quasilocal Corrections to Bondi's Mass-Loss Formula and Dynamical
Horizons}
\author{Albert Huber\thanks{hubera@technikum-wien.at}}
\date{{\footnotesize{}UAS Technikum Wien - Department of Applied Mathematics
		and Physics, H{\"o}chst{\"a}dtplatz 6, 1200 Vienna, Austria}}
\maketitle
\begin{abstract}
In this work, a null geometric approach to the Brown-York quasilocal
formalism is used to derive an integral law that describes the rate
of change of mass and/or radiative energy escaping through a dynamical
horizon of a non-stationary spacetime. The result thus obtained shows
- in accordance with previous results from the theory of dynamical
horizons of Ashtekar et al. - that the rate at which energy is transferred
from the bulk to the boundary of spacetime through the dynamical horizon
becomes zero at equilibrium, where said horizon becomes non-expanding
and null. Moreover, it reveals previously unrecognized quasilocal
corrections to the Bondi mass-loss formula arising from the combined
variation of bulk and boundary components of the Brown-York Hamiltonian,
given in terms of a bulk-to-boundary inflow term akin to an expression
derived in an earlier paper by the author \cite{huber2022remark}.
For clarity, this is discussed with reference to the Generalized Vaidya
family of spacetimes, for which derived integral expressions take
a particularly simple form.
\end{abstract}
\textit{\footnotesize{}Key words: quasilocal Hamiltonian, dynamical
horizons, Bondi mass-loss formula}{\footnotesize\par}

\section*{Introduction}

To determine within the Brown-York quasilocal formalism \cite{booth1999moving,brown1993quasilocal}
the change in mass and/or radiant energy escaping through the spatial
boundary of a finitely extended gravitating physical system, it generally
incurs, as only recently shown in \cite{huber2022remark}, the necessity
to calculate the time derivative of the total quasilocal gravitational
Hamiltonian (bulk plus boundary term) rather than just that of the
boundary part. The main reason for this is that the temporal variation
of the ADM Hamiltonian, which corresponds to the bulk part of the
total expression mentioned above, yields a non-vanishing bulk-to-boundary
inflow term that leads to corrections to Einstein's quadrupole formula
in the linearized weak-field approximation of general relativity.

This integral term, if different from zero (which is possible only
in the non-vacuum case), has been shown to play a role in the quasilocal
description of various physical phenomena, such as tidal deformation
and heating processes as well as gravitoelectromagnetic effects \cite{huber2022remark}.
Moreover, as has also been shown, its existence entails some remarkable
consequences, perhaps the most striking of which is that the corrections
it causes lead to a shift in the overall intensity of gravitational
radiation emanating from compact gravitational sources such as stars
and black holes. This is remarkable not least because the intensity
shift in question should in principle prove to be experimentally detectable
resp. observable in gravitational wave simulations, thus leading to
a physical prediction that can readily be tested with modern methods
of gravitational wave astronomy.

The main problem in this respect, however, is that it has not yet
been clearly established whether the corrections caused by the mentioned
inflow term are smaller or of the same order of magnitude as other
integral terms resulting from the variation of the quasilocal Brown-York
Hamiltonian. Moreover, with the exception of selected models of linearized
Einstein-Hilbert gravity, the precise physical meaning of the corrections
in question has remained elusive to this day.

In response to these shortcomings, the present work takes a specific
approach to the subject by calculating within a bounded non-stationary
spacetime the flux of mass and/or radiant energy through the dynamical
horizon of the geometry, as well as its temporal variation. As a basis
for these calculations, a null geometric approach to the quasilocal
Brown-York formalism is pursued, which is shown to be compatible with
the powerful dynamical horizon framework of Ashtekar et al. \cite{ashtekar2003dynamical,ashtekar2004isolated}
and Hayward's related trapping horizon approach \cite{hayward1994general}.
To this end, following a previous work on the subject \cite{booth2005horizon},
a geometric setting is introduced that involves a spatially and temporally
bounded spacetime with inner and outer boundaries, where the inner
boundary is given by a dynamical horizon. Regarding this particular
geometric setting, the time-flow vector field of spacetime is then
chosen to coincide once with the lightlike horizon vector field of
the geometry (which is generally non-tangential to said horizon) and
once with the same horizon vector field plus a boundary shift vector,
and the resulting total Hamiltonian is varied with respect to these
same vector fields. Thereby, it is shown that the methods used naturally
lead to a null geometric equivalent of the bulk-to-boundary inflow
term derived in \cite{huber2022remark} and thus to a corresponding
intensity shift of emitted gravitational radiation.

The latter is concluded from the fact that the resulting quasilocal
corrections do not vanish even if the outer boundary of spacetime
is shifted to infinity in the large sphere limit. In lieu thereof,
as shown in the second section of the paper, a modification of Bondi's
celebrated mass-loss formula \cite{bondi1962gravitational,Madler:2016xju}
results in such a case, which shows that radiative contributions at
infinity can occur even if the Bondi news function is zero, and thus
supposedly the time derivative of the associated mass aspect. It thus
appears that, according to the quasilocal Hamiltonian formalism, there
are exceptions to the generally accepted rule: \textit{The mass of
a system is constant if and only if there is no news}. As it seems,
no similar result has been obtained in the literature so far. The
quasilocal corrections responsible for this fact are determined explicitly
in section two of the work. 

For the standard choice for the lapse function proposed in the dynamical
horizon framework, the result thus obtained shows that the temporal
variation of the total quasilocal Brown-York Hamiltonian vanishes
once the horizon reaches a steady state of equilibrium and becomes
an isolated or weakly isolated horizon in the sense of \cite{ashtekar2000generic,ashtekar1999isolated,ashtekar2000mechanics,ashtekar2002geometry,ashtekar2000isolated}.
Thus, in agreement with the common expectation, the discussed model
confirms that any matter and/or radiation flux (of the specified type)
from the bulk to the boundary of spacetime that crosses a dynamical
horizon necessarily subsides completely in the limiting case where
the local horizon geometry becomes stationary and settles into a state
of equilibrium, as in the case of a black hole.

To eventually assess the magnitude of the integral terms involved
and to provide an explicit example of non-vanishing radiative contributions
at infinity, the corresponding expressions are calculated in the third
and final section of the paper with respect to models of the Generalized
Vaidya spacetime family, for which the resulting integral expressions
take a particularly simple form in case that the boundary of spacetime
is shifted to infinity in the large sphere limit. In doing so, it
is shown $i)$ that radiation fields may be detected at null infinity
even in cases where the Bondi news function is zero, and $ii)$ that
the resulting quasilocal corrections depend to a large extent on the
choice of the time-flow vector field of the geometry. Potential implications
of these findings are discussed towards the end of the paper.

\section{Quasilocal Hamiltonian and Mass-Energy Transfer in Bounded Gravitational
Fields}

In this first preliminary section, the geometric setting to be considered
is introduced, and some of the main results of \cite{huber2022remark}
are recapitulated and generalized to fit this same setting. In particular,
the time derivative of the quasilocal Brown-York gravitational Hamiltonian
is calculated in a spacetime with interior and exterior boundaries,
leading to an integral law describing how the matter and/or radiation
content of a spatially and temporally bounded gravitating physical
system changes with time. The bulk-to-boundary inflow term mentioned
in the introduction is derived in the process, and it is shown what
form some of the relevant integral expressions take with respect to
the special choice of a lightlike (horizon) time-flow vector field
of spacetime.

As a basis for the ensuing calculations, the present section essentially
takes up the geometric setting considered in \cite{huber2022remark}.
However, the latter is to be extended to comply with the dynamical
horizon framework of Ashtekar et al. \cite{ashtekar2003dynamical}
in a manner similar to an earlier, slightly related work by Booth
and Fairhurst \cite{booth2005horizon}. For this purpose, let a fully
dynamical, spatially compact, time orientable spacetime $(\mathcal{M},g)$
with manifold structure $\mathcal{M}\equiv M\cup\partial M$ be considered,
which is foliated by a family of $t=const.$-hypersurfaces $\{\Sigma_{t}\}$.
This spacetime may be envisioned as a non-stationary spacetime in
a 'box', i.e., a dynamical spacetime with a cylindrical outer boundary,
the latter being later shifted to infinity. In more concrete terms,
the boundary $\partial M$ of said spacetime shall consist of two
parts: an exterior part $\partial M_{ext}$ and an an interior part
$\partial M_{int}$ such that $\partial M\equiv\partial M_{int}\cup\partial M_{ext}$.
The exterior part $\partial M_{ext}$ of the boundary shall be chosen
in such a way that $\partial M_{ext}\equiv\Sigma_{1}\cup\mathcal{B}\cup\Sigma_{2}$
applies, where $\Sigma_{1}$ and $\Sigma_{2}$ represent spatial boundary
parts, while $\mathcal{B}$ represents a timelike boundary portion.
This timelike portion shall be assumed to be foliated by a collection
of two-surfaces $\{\Omega_{t}\}$ such that $\mathcal{B}=\{\underset{t}{\cup}\Omega_{t}:\,t_{1}\leq t\leq t_{2}\}$.
Additionally, it shall be assumed that there exists an interior boundary
$\partial M_{int}\equiv\mathcal{S}_{1}\cup\mathcal{T}\cup\mathcal{S}_{2}$,
where $\mathcal{T}$ is a spacelike hypersurface representing a (canonical)
dynamical horizon in the sense of Ashtekar et al. That is to say,
$\mathcal{T}$ is assumed to be a smooth, three-dimensional, spacelike
submanifold of spacetime that exhibits a foliation $\{\mathcal{S}_{t}\}$
by marginally trapped surfaces such that relative to each leaf of
the foliation there exist two null normals $l^{a}$ and $k^{a}$ and
two associated null expansion scalars $\Theta=q^{ab}\nabla_{a}l_{b}$
and $\Xi=q^{ab}\nabla_{a}k_{b}$, where $q^{ab}$ is the inverse of
the induced metric $q_{ab}=g_{ab}+l_{a}k_{b}+k_{a}l_{b}$, one of
which vanishes locally and the other of which is strictly negative,
i.e. $\Theta=0$ and $\Xi<0$ on $\mathcal{T}$. 

\begin{figure}
\includegraphics[viewport=-275bp 0bp 417bp 570.5bp,scale=0.35]{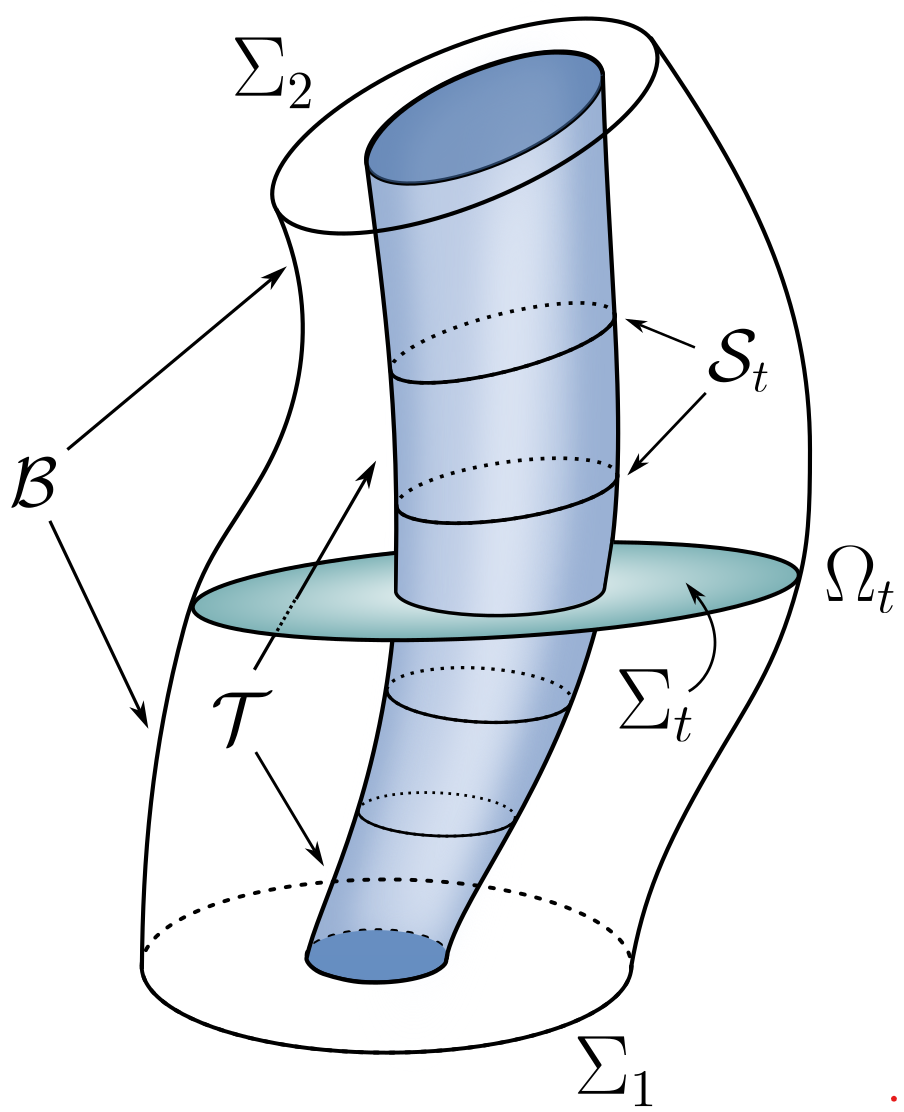}\caption{A schematic three-dimensional representation of the spacetime manifold
$M$ along with its boundaries.}
\end{figure}

Taking these assumptions as a starting point, the results of \cite{huber2022remark}
shall be recapitulated in the following. To this end, the conventions
of the mentioned work shall be adopted and adapted to the given geometric
setting. To start with, a future-directed time evolution vector field
$t^{a}=Nn^{a}+N^{a}$ shall be considered, which, at the timelike
boundary $\mathcal{B}$, takes the form $t^{a}=\mathcal{N}v^{a}+\mathcal{N}^{a}$,
where $N$ and $N^{a}$ are the corresponding lapse function and shift
vector field as usual, $n^{a}$ is the normalized timelike generator
leading to the the spacelike slicing of $(\mathcal{M},g)$, $v^{a}$
is some timelike vector field tangent to $\mathcal{B}$ and orthogonal
to $\Omega_{t}$ and $\mathcal{N}$ and $\mathcal{N}^{a}$ are the
corresponding boundary lapse function and boundary shift vector field,
respectively. Given this vector field and the related conventions,
the corresponding three-metric at $\Sigma_{t}$ reads $h_{ab}=g_{ab}+n_{a}n_{b}$.
To define the induced three-metric $\gamma_{ab}$ at $\mathcal{B}$,
on the other hand, one may additionally consider a spatial unit vector
field $u^{a}$ which is perpendicular to $\mathcal{B}$ and thus orthogonal
to the temporal unit vector field $v^{a}$ tangent to $\mathcal{B}$.
With respect to the latter, the three-metric then is $\gamma_{ab}=g_{ab}-u_{a}u_{b}$.
Moreover, considering a further spacelike vector field $s^{a}$ that
is orthogonal to the timelike generator $n^{a}$ of the spacelike
foliation $\{\Sigma_{t}\}$, but generally non-orthogonal to $v^{a}$
(in contrast to $u^{a}$, which is generally non-orthogonal to $n^{a}$),
one finds that the induced two-metric $q_{ab}$ at $\Omega_{t}$ takes
the form $q_{ab}=g_{ab}-u_{a}u_{b}+v_{a}v_{b}=g_{ab}+n_{a}n_{b}-s_{a}s_{b}$.
Using the latter relation in combination with the decompositions $n^{a}=\frac{1}{\sqrt{2}}(l^{a}+k^{a})$
and $s^{a}=\frac{1}{\sqrt{2}}(l^{a}-k^{a})$ of $n^{a}$ and $s^{a}$,
where $l^{a}$ and $k^{a}$ are null normals reducing locally to those
associated with a given leaf $\mathcal{S}_{t}$ of the foliation $\{\mathcal{S}_{t}\}$
of the dynamical horizon $\mathcal{T}$, it then becomes clear that
the induced metric at said horizon takes the previously claimed form
$q_{ab}=g_{ab}+l_{a}k_{b}+k_{a}l_{b}$. With respect to this induced
metric, the boundary shift vector can be written in the form $\mathcal{N}^{a}=q_{\;c}^{a}N^{c}$.

The consideration of all the foregoing definitions proves to be beneficial
for setting up the quasilocal Brown-York-Hamiltonian 
\begin{equation}
H=H_{0}+H_{h},
\end{equation}
which in the given context consists of two parts $H_{0}$ and $H_{h}$,
both of which themselves consist of bulk and boundary parts such that 

\begin{equation}
H_{0}=H_{0}^{Bulk}+H_{0}^{Boundary};\quad H_{h}=H_{h}^{Bulk}+H_{h}^{Boundary},
\end{equation}
where 
\begin{align}
H_{0}^{Bulk}= & \underset{\Sigma_{t}}{\int}\mathscr{H}d^{3}x,\;H_{0}^{Boundary}=\underset{\Omega_{t}}{\int}\mathcal{\mathfrak{H}}d^{2}x\\
H_{h}^{Bulk}= & \underset{\mathcal{T}}{\int}\mathscr{H}d^{3}x,\;H_{h}^{Boundary}=\underset{\mathcal{S}_{t}}{\int}\mathcal{\mathfrak{H}}d^{2}x\nonumber 
\end{align}
shall apply by definition. Here, $\mathscr{H}$ represents the ADM
Hamiltonian density $\mathscr{H}=\frac{\sqrt{h}}{8\pi}G_{ab}t^{a}n^{b}=\frac{\sqrt{h}}{8\pi}(N\mathcal{H}+\mathcal{H}_{a}N^{a})$.
The individual parts of this Hamiltonian density read $\mathcal{H}=\frac{1}{2}[K_{ab}K^{ab}-K^{2}+{}^{(3)}R]$
and $\mathcal{H}{}_{a}=D_{b}(K_{\;a}^{b}-h_{\;a}^{b}K)$, where $K_{ab}=\frac{1}{2}h_{\;a}^{c}h_{\;b}^{d}L_{n}h_{cd}=\frac{1}{2N}(\dot{h}_{ab}+2D_{(a}N_{b)})$
is the extrinsic curvature of the three-hypersurface $\Sigma_{t}$
and $D_{a}$ is the covariant derviative at $\Sigma_{t}$. The Hamiltonian
density $\mathcal{\mathfrak{H}}$ at $\Omega_{t}$, on the other hand,
can be written in the form $\mathcal{\mathfrak{H}}=\frac{\sqrt{q}}{8\pi}\rho_{ab}t^{a}v^{b}=\frac{\sqrt{q}}{8\pi}(\mathcal{N}\mathfrak{h}+\mathfrak{h}_{a}\mathcal{N}^{a})$,
by virtue of the fact that $\mathfrak{h}=\frac{\sqrt{q}}{8\pi}\mathfrak{K}$
and $\mathfrak{h}_{a}=\frac{\sqrt{q}}{8\pi}u^{b}\mathcal{D}_{a}v_{b}$
with $\mathfrak{K}=q^{ab}\mathfrak{K}_{ab}=q^{ab}\mathcal{D}_{a}u_{b}$
applies in this context, where $\mathfrak{K}_{ab}$ is the extrinsic
curvature calculated with respect to $u^{a}$ and $\mathcal{D}_{a}$
the covariant derviative at $\Omega_{t}$. Using the well-known decomposition
relations $v^{a}=\frac{1}{\lambda}n^{a}-\eta s^{a}$ and $u^{a}=\frac{1}{\lambda}s^{a}-\eta n^{a}$,
given in terms of the parameter $\eta=u_{a}n^{a}=-s_{a}v^{a}$ measuring
the non-othogonality of $n^{a}$ and $u^{a}$ as well as $v^{a}$
and $s^{a}$ and a related boost parameter $\lambda=\frac{1}{\sqrt{1+\eta^{2}}}$,
one finds here that $\mathcal{\mathfrak{H}}$ can alternatively be
specified in terms of $n^{a}$ and $s^{a}$. In particular, the identity 

\begin{equation}
\mathcal{N}\mathfrak{K}=Nk-(K_{ab}-Kh_{ab})N^{a}s^{b}-\lambda(\mathcal{N}\mathcal{D})\eta-(\mathcal{N}\mathcal{D})v_{a}u^{a}
\end{equation}
is found to be valid in this context, which will play a role later
in determining changes in the matter and/or radiation content of the
system at infinity as well as through the dynamical horizon $\mathcal{T}$.
The quantity $k=q^{ab}k_{ab}=q^{ab}\mathcal{D}_{a}s_{b}$ entering
this very identity is admittedly the extrinsic curvature of $\Omega_{t}$,
calculated with respect to $s^{a}$. 

As a supplementary remark, it may be noted that $\mathcal{H}$ and
$\mathcal{H}_{a}$ as well as $\mathfrak{h}$ and $\mathfrak{h}_{a}$
could be specified completely analogously at $\mathcal{T}$ and $\mathcal{S}_{t}$
in terms of the corresponding conjugate momenta (even if only with
respect to the first and second fundamental forms of said hypersurface
and surface, respectively). However, taking such a step will not prove
necessary in the following and will therefore not be undertaken at
this point. Instead, direct recourse will be made to the null geometric
framework used in \cite{ashtekar2003dynamical} in the second section
of the paper, rendering a specification of said quantities obsolete.

That said, the next step will be to calculate the time derivative
$\frac{dH}{dt}$ of the quasilocal Hamiltonian by performing a variation
$\mathsf{\mathfrak{L}}_{t}$ of each of the integral expressions in
$(3)$, where $\mathsf{\mathfrak{L}}_{t}$ denotes the Lie derivative
along the spacelike section $\Sigma_{t}$ with respect to the time
evolution vector $t^{a}$. This yields a result of the form $\frac{dH}{dt}=\frac{dH_{0}}{dt}+\frac{dH_{h}}{dt}=\mathsf{\mathfrak{L}}_{t}H_{0}+\mathsf{\mathfrak{L}}_{t}H_{h}$. 

The emphasis is here first put on the calculation of the former term
$\frac{dH_{0}}{dt}$ of the time derivative of $H$ and only then
on that of the second term $\frac{dH_{h}}{dt}$, which proceeds completely
analogously. For the purpose of calculating said term, one may take
into account that expressions split once more in bulk and boundary
parts, i.e. $\frac{dH_{0}}{dt}=\frac{dH_{0}^{Bulk}}{dt}+\frac{dH_{0}^{Boundary}}{dt}$,
and use the identity 

\begin{align}
\mathsf{\mathfrak{L}}_{t}\mathcal{H}+\mathsf{\mathfrak{L}}_{t}\mathcal{H}_{a}\cdot N^{a} & =\frac{N}{2}\mathcal{Q}_{ab}\mathsf{\mathfrak{L}}_{t}h^{ab}-(NK+D_{a}N^{a})(N\mathcal{H}+\mathcal{H}_{b}N^{b})+\\
 & +D_{a}[(N\mathcal{H}+\mathcal{H}_{b}N^{b})N^{a}+N^{2}\mathcal{H}^{a}+2N\mathcal{Q}{}_{\;b}^{a}N^{b}],\nonumber 
\end{align}
in combination with $\mathsf{\mathfrak{L}}_{t}\sqrt{h}=\sqrt{h}(NK+D_{a}N^{a})$
to obtain the integral expressions

\begin{equation}
\frac{dH_{0}^{Bulk}}{dt}=\underset{\Sigma_{t}}{\int}(\dot{N}\mathcal{H}+\mathcal{H}_{a}\dot{N}^{a}+\frac{N}{2}\mathcal{Q}_{ab}\dot{h}^{ab})\omega_{h}+\underset{\Omega_{t}}{\int}\Pi\omega_{q},
\end{equation}
and 
\begin{equation}
\frac{dH_{0}^{Boundary}}{dt}=\underset{\Omega_{t}}{\int}(\dot{\mathcal{N}}\mathfrak{h}+\mathfrak{h}_{a}\dot{\mathcal{N}}^{a}+\frac{\mathcal{N}}{2}\mathfrak{Q}_{ab}\dot{q}^{ab})\omega_{q},
\end{equation}
where
\begin{align}
\mathcal{Q}_{ab} & =\frac{\sqrt{h}}{8\pi}\{{}^{(3)}R_{ab}-2K_{ac}K_{\;b}^{c}+KK_{ab}-\frac{1}{N}\left[\dot{K}_{ab}+(ND)K_{ab}+D_{a}D_{b}N\right]-\\
 & -\frac{1}{2}h_{ab}\left(^{(3)}R+K^{2}-K_{cd}K^{cd}-\frac{2}{N}\left[\dot{K}+(ND)K+D_{a}D^{a}N\right]\right)\};\nonumber \\
\mathfrak{Q}_{ab} & =\frac{\sqrt{q}}{8\pi}(\mathfrak{K}_{ab}-(\mathfrak{K}-b_{a}u^{a})q_{ab})\nonumber 
\end{align}
applies by definition. To obtain the above integral expressions, as
should be noted, the definitions $\mathcal{Q}_{ab}=h_{a}^{\;c}h_{b}^{\;d}G_{cd}$,
$\mathfrak{Q}_{ab}=q_{a}^{\:c}q_{b}^{\:d}\rho_{cd}$, $b^{a}\equiv(v\nabla)v^{a}$,
$\dot{N}=\mathsf{\mathfrak{L}}_{t}N$, $\dot{N}^{a}=\mathsf{\mathfrak{L}}_{t}N^{a}$,
$\dot{h}^{ab}=\mathsf{\mathfrak{L}}_{t}h^{ab}$ as well as $\dot{\mathcal{N}}=\mathcal{L}_{t}\mathcal{N}$,
$\dot{\mathcal{N}}^{a}=\mathcal{L}_{t}\mathcal{N}^{a}$, $\dot{q}^{ab}=\mathcal{L}_{t}q^{ab}$
have been used, where $\mathcal{L}_{t}$ denotes the induced Lie-derivative
at $\Omega_{t}$ pointing along $t^{a}$. 

As can be seen, the result thus obtained now decomposes into three
terms: a bulk term, a boundary term and the bulk-to-boundary inflow
term already mentioned in the introduction. The latter term occurring
in $(6)$ results from a total divergence and is given with respect
to an integrand of the form
\begin{equation}
\Pi=\frac{1}{8\pi}[(N\mathcal{H}+\mathcal{H}_{b}N^{b})N_{a}s^{a}+N^{2}\mathcal{H}_{a}s^{a}+N\mathcal{Q}_{ab}N^{a}s^{b}].
\end{equation}
From this, it is found that relations $(2)$, $(3)$, $(6)$ and $(7)$
give rise to a power functional of the form

\begin{equation}
\mathscr{P}_{0}=\frac{dH_{0}^{Boundary}}{dt}+\underset{\Omega_{t}}{\int}\Pi\omega_{q}=\underset{\Omega_{t}}{\int}\mathcal{I}\omega_{q},
\end{equation}
where the occurring intensity expression $\mathcal{I}$ reads
\begin{equation}
\mathcal{I}=\mathcal{I}_{0}+\Pi
\end{equation}
with $\mathcal{I}_{0}:=\dot{\mathcal{N}}\mathfrak{h}+\mathfrak{h}_{a}\dot{\mathcal{N}}^{a}+\frac{\mathcal{N}}{2}\mathfrak{Q}_{ab}\dot{q}^{ab}$
and $\frac{dH_{0}^{Boundary}}{dt}=\mathcal{L}_{t}H_{0}^{Boundary}$.
The default candidate $\mathcal{I}_{0}$ for such an intensity, previously
derived in \cite{booth1999moving,booth2000quasilocal}, is therefore
shifted by a $\Pi$-term of the form $(9)$, which is zero only in
the vacuum case. In all other cases, this term is generally different
from zero, which implies that the corresponding surface integral in
$(6)$ and $(10)$ does not vanish even if the outer boundary of spacetime
is shifted to infinity in the large sphere limit; a point from which
it was concluded in \cite{huber2022remark} that quasilocal corrections
of Einstein's quadrupole formula arise in the same limit.

This, of course, applies to any choice of the time-flow vector field
$t^{a}$. In particular, it applies to a class of such vector fields
arising from an orthogonal $2+1$-decomposition $N^{a}=Os^{a}+\mathcal{N}^{a}$
of the shift vector with respect to the surface $\Omega_{t}$, which
yields a bulk-to-boundary inflow term with an integrand of the form

\begin{equation}
\Pi=\Pi_{0}+\varPi_{N};
\end{equation}
by virtue of the fact that the definitions $\Pi_{0}:=\frac{N^{2}}{8\pi}\mathcal{H}_{a}s^{a}$
and $\varPi_{N}:=\frac{1}{8\pi}[O(N\mathcal{H}+O\mathcal{H}_{a}s^{a}+\mathcal{H}_{a}\mathcal{N}^{a})+O\cdot N\mathcal{Q}_{ab}s^{a}s^{b}+N\mathcal{Q}_{ab}s^{a}\mathcal{N}^{b}]$
are used in this context.

As may be noted, the foreging results can be generalized in the sense
that one may choose a linear combination of the form $\xi^{a}=t^{a}+\Omega\varphi^{a}$
as time evolution vector field of spacetime, where $\varphi^{a}$
is an angular vector field tangential to all the cross-sections of
$\Sigma_{t}$, i.e. a vector field that coincides with a corresponding
Killing field in the regime in which the dynamical horizon framework
tends to the isolated horizon framework; a regime in which spacetime
typically exhibits global generators of time translations and rotations.
In this case, the form of the corresponding quasilocal corrections
can be straightforwardly determined from the above, using the fact
that the vector field $\xi^{a}$ can be decomposed in the form $\xi^{a}=\tilde{N}n^{a}+\tilde{N}^{a}$
with $\tilde{N}=N-\Omega n_{b}\varphi^{b}$ and $\tilde{N}^{a}=N^{a}+h_{\:b}^{a}\varphi^{b}=\tilde{O}s^{a}+\tilde{\mathcal{N}}^{a}$
and making the replacements $N\rightarrow\tilde{N}$, $O\rightarrow\tilde{O}$
and $\mathcal{N}^{a}\rightarrow\tilde{\mathcal{N}}^{a}$ in $(12)$,
which yields the analogous expression 

\begin{equation}
\tilde{\Pi}=\tilde{\Pi}_{0}+\tilde{\varPi}_{\tilde{N}}
\end{equation}
with $\tilde{\Pi}_{0}:=\tilde{N}^{2}\mathcal{H}_{a}s^{a}$. Accordingly,
the problem of determining quasilocal corrections in the rotating
case can be handled exactly along the same lines as in the non-rotating
case; with the only difference being that $\tilde{N}$ and $\tilde{N}^{a}$
are now the associated shifted versions of the lapse function and
the shift vector field of the spacetime metric. Otherwise, there is
no difference in the treatment of these cases.

That said, let it be noted that there is an important special case
resulting from $(12)$, which arises when the time-flow vector of
the geometry is chosen to be $t^{a}=\sqrt{2}Nl^{a}$, with $l^{a}=\frac{1}{\sqrt{2}}(n^{a}+s^{a})$
being a null vector field that reduces locally to the horizon vector
field of spacetime at $\mathcal{T}$. The latter follows directly
from $(9)$ for the case that $O\equiv N$, $\mathcal{N}^{a}\equiv0$
and thus $N^{a}=Ns^{a}$ is chosen to be satisfied. Given precisely
this choice, the identity $\mathcal{H}+2\mathcal{H}_{a}s^{a}+\mathcal{Q}_{ab}s^{a}s^{b}=G_{ab}n^{a}n^{b}+2G_{ab}n^{a}s^{b}+G_{ab}s^{a}s^{b}=2G_{ab}l^{a}l^{b}$
can be used to convert the integrand of the bulk-to-boundary integral
term in $(6)$, leading to the result

\begin{equation}
\Pi=\varPi=\frac{N^{2}}{4\pi}G_{ab}l^{a}l^{b}.
\end{equation}
Moreover, the further identity $K-K_{ab}s^{a}s^{b}+k=\sqrt{2}\Theta$
can be used to convert the boundary Hamiltonian density $\mathcal{\mathfrak{H}}$
into
\begin{equation}
\mathcal{\mathfrak{H}}=\frac{\sqrt{2q}N\Theta}{8\pi}.
\end{equation}
This makes it clear that the boundary Hamiltonian vanishes for the
given choice of time-flow vector field whenever $\Theta=0$ applies
at the exterior spatial boundary $\mathcal{B}$ of spacetime (that
is, in particular, when said boundary constitutes a timelike dynamical
horizon); quite in contrast to the temporal variation of said term
with respect to the null time evolution vector field $t^{a}=\sqrt{2}Nl^{a}$,
which is generally different from zero.

As will be shown in the subsequent section of this work, the latter
proves to be particularly important in that said variation of the
boundary Hamiltonian - after being combined with a bulk-to-boundary
inflow term with an integrand of the form $(14)$ - leads to Bondi's
result for mass loss in gravitating systems due to gravitational radiation;
while other choices typically lead to quasilocal corrections to Bondi's
formula. This is the reason why, in order to capture deviations from
Bondi's mass-loss formula and simultaneously quantify the strength
of the aforementioned quasilocal corrections, it will prove useful
in the following to make an ansatz of the form $t^{a}=\sqrt{2}Nl^{a}+V^{a}$
for the time evolution vector of spacetime, where $V^{a}=-Ns^{a}+N^{a}=(O-N)s^{a}+\mathcal{N}^{a}$
must be satisfied for the sake of consistency. 

Provided that this is indeed the case, the ansatz mentioned proves
to be fully compatible with all foregoing results, yielding a $\Pi$-term
of the form $(12)$ and a boundary Hamiltonian density given by the
expression

\begin{equation}
\mathcal{\mathfrak{H}}=\frac{\sqrt{q}}{8\pi}\left[\sqrt{2}N\Theta-\varGamma_{V}\right],
\end{equation}
where the quantity $\varGamma_{V}:=(K_{ab}-Kh_{ab})s^{a}V^{b}+\lambda(\mathcal{N}\mathcal{D})\eta$
has been introduced; a quantity that proves consistent with that of
the null Brown-York tensor given in \cite{chandrasekaran2022brown}
for $O=\eta=0$ and $\Omega_{a}=q_{\;a}^{c}K_{bc}s^{b}$, where $\Omega_{a}=-q_{a}^{\;b}k^{c}\nabla_{b}l_{c}$
is the Haji$\hat{c}$ek one-form. 

As may be noted, in order to return to the case of a rotating black
hole, just the replacements $N\rightarrow\tilde{N}$, $O\rightarrow\tilde{O}$,
and $\mathcal{N}^{a}\rightarrow\tilde{\mathcal{N}}^{a}$ need to be
made in this context. Such a transition has the interesting consequence
that part of the boundary component $H_{0}^{Boundary}$ of the quasilocal
Hamiltonian gives rise to an integral expression of the form
\begin{equation}
J_{0}^{\Omega\varphi}=\frac{1}{8\pi}\underset{\Omega_{t}}{\int}[\Omega(K_{ab}-Kh_{ab})s^{a}\varphi^{b}]\omega_{q},
\end{equation}
which can be converted to agree with a generalized version of Komar's
angular momentum integral by applying Gauss' theorem and using the
momentum constraint equation. Hence, as a direct consequence, it is
found that the exterior boundary part of the Hamiltonian splits up
in two parts, i.e. $H_{0}^{Boundary}=H_{0,red}^{Boundary}-J_{0}^{\Omega\varphi}$,
where $J_{0}^{\Omega\varphi}$coincides with the ADM angular momentum
associated with $\Omega\varphi^{b}$. Accordingly, the variation of
$J_{0}^{\Omega\varphi}$ with respect to $t^{a}$ necessarily yields
a torque term $M_{0}^{\Omega\varphi}\equiv\frac{dJ_{0}^{\Omega\varphi}}{dt}$
which fully characterizes the power of the rotational motion of the
system at the boundary of spacetime, especially for spacetimes with
asymptotic rotational symmetry when the latter is shifted to infinity.

With this now being clarified, it may next be noted that, for all
the foregoing applies in a similar way to fields at the interior boundary
of spacetime, the variation of the horizon part $H_{h}$ of the Hamiltonian
$H$ can be calculated in exactly the same way as above. Yet, since
$\mathcal{T}$ is a dynamical horizon in the sense of Ashtekar et
al. it is clear that when $t^{a}=\sqrt{2}Nl^{a}$ is chosen to be
the time-flow vector of spacetime, the boundary part of this part
of the quasilocal Hamiltonian will generally be zero, while its variation
will in turn generally be different from zero. Therefore, also in
the given case, a quasilocal power functional of the form

\begin{equation}
\mathscr{P}_{h}=\frac{dH_{h}^{Boundary}}{dt}+\underset{\mathcal{S}_{t}}{\int}\Pi\omega_{q}=\underset{\mathcal{S}_{t}}{\int}\mathcal{I}\omega_{q}
\end{equation}
can be defined, where $\frac{dH_{h}^{Boundary}}{dt}=\mathcal{L}_{t}H_{h}^{Boundary}$
holds by definition and the $\Pi$-term is given by $(14)$. 

In this case, too, it proves useful to consider the shifted horizon
vector field $t^{a}=\sqrt{2}Nl^{a}+V^{a}$, if only to to be able
to include again the rotating case by choosing $V^{a}=\Omega\varphi^{a}$,
where in case of a black hole spacetime $\Omega$ and $\varphi^{a}$
represent the angular velocity of the black hole and an associated
angular vector field coincident with the Killing field of the geometry
when the black hole spacetime under consideration is axisymmetric.
This choice is intriguing not least because it allows one to derive
(with respect to a portion $\Delta\mathcal{T}$ of the dynamical horizon
$\mathcal{T}$) the Ashtekar-Krishnan version of the first law of
black hole mechanics from \cite{ashtekar2003dynamical}; a law which
- similar to Hayward's first law of black hole dynamics \cite{hayward1994general},
but different from the original law of Bardeen, Carter and Hawking
\cite{bardeen1973four} - remains valid even in the light of dynamical
black hole spacetimes. Still, a more generic scenario arises in the
given context if again simply the replacements $N\rightarrow\tilde{N}$,
$O\rightarrow\tilde{O}$ and $\mathcal{N}^{a}\rightarrow\tilde{\mathcal{N}}^{a}$
are made, yielding in full analogy to the above a splitting $H_{h}^{Boundary}=H_{h,red}^{Boundary}-J_{h}^{\Omega\varphi}$
of the boundary Hamiltonian, where the corresponding horizon angular
momentum $J_{h}^{\Omega\varphi}$ is of the exact same form as $J_{0}^{\Omega\varphi}$
depicted in $(17)$, except that the latter is defined with respect
to the cut $\Omega_{t}$, while the former is defined with respect
to $\mathcal{S}_{t}$. The mentioned horizon angular momentum then
leads again to a torque term $M_{h}^{\Omega\varphi}\equiv\frac{dJ_{h}^{\Omega\varphi}}{dt}$,
which characterizes the power of the rotational motion of the system
along the horizon. 

This in advance, it may next be noted that a large part of the upcoming
section will be devoted to a more detailed characterization of equations
$(10)$ and $(18)$, for which purpose a null geometric derivation
of the surface integrals with integrands of the form $(12)$ and $(13)$
will be given and the derivatives $\frac{dH_{h}^{Boundary}}{dt}$
and $\frac{dH_{0}^{Boundary}}{dt}$ of the boundary parts of the Hamiltonian
will be calculated with regard to the shifted horizon vector field
$t^{a}=\sqrt{2}Nl^{a}+V^{a}$; first for $V^{a}=0$ and then for $V^{a}\neq0$.
The results obtained in this way pass an important test along the
way in that they are found to be fully consistent with the theory
of dynamical and isolated horizons of Ashtekar et al. Moreover, it
is found that the interpretation of the surface integrals over $\Pi$
and $\tilde{\Pi}$ as bulk-to-boundary inflow terms also proves to
be absolutely tenable, not least because - given a suitable choice
of the lapse function at the horizon - it can be observed that the
rate of energy transfer from the bulk through the inner to the outer
boundary of spacetime (and vice versa) becomes zero in the limiting
case where the dynamical horizon of spacetime transitions to an isolated
or weakly isolated horizon and settles into a stable equilibrium state;
a state in which it would be impossible for matter to cross the outermost,
non-expanding null horizon and then escape to infinity, as in the
case of a black hole. The approach taken in this paper thus reflects
this particular aspect of black hole physics, as it should if an interpretation
of the derived integral expressions as bulk-to-boundary inflow terms
were to prove plausible. The latter is further clarified in the third
and final section of this work by the concrete example of the Generalized
Vaidya family of spacetimes.

\section{Matter and Radiation Transfer through Dynamical Horizons}

Having obtained a number of results applicable to quantities at both
the inner and outer boundaries of spacetime, the present section is
now devoted to the calculation of exactly the same quantities from
a different angle; thus continuing the quasilocal description of mass
and radiative energy transfer in bounded non-stationary spacetimes
with dynamical horizons begun in the previous section. 

For the calculation of these quantities, a null geometric approach
is adopted this time, by which it is shown that the results deduced
in the previous section prove to be consistent with, and are derivable
within, the theory of dynamical horizons. Furthermore, it is shown
that the corresponding boundary terms - depending on the choice of
the time-flow vector field of the geometry - either reduce directly
to the Bondi mass-loss formula or lead to quasilocal corrections from
the latter when the outer boundary of spacetime is shifted to infinity.

As a first step in dealing with the above and thus linking the results
of the previous section to the theory of dynamical horizons of Ashtekar
et al., let a radial parameter $R$ be considered and used as a coordinate
for describing local effects at the horizon $\mathcal{T}$. Given
this null coordinate, the choice $N\equiv N_{R}$ can be made for
the lapse function, where $N_{R}\equiv\vert\partial R\vert$ shall
by definition apply in this context. This choice for the lapse, as
described in detail in the relevant literature on the subject, turns
out to be favorable for several reasons; one of which is that it causes
the ADM Hamiltonian $H_{h}^{bulk}$ on the black hole horizon $\mathcal{T}$
(and, in fact, the entire horizon part $H_{h}$ of the Brown-York
Hamiltonian $H$) to vanish as soon as the latter transitions from
a dynamical to an isolated or weakly isolated horizon, thereby ensuring
that the rate of transferred energy becomes zero once the geometry
of the black hole becomes stationary and its horizon non-expanding
and null. On top of that, some of the ensuing calculations can be
greatly simplified by choosing the lapse function in this particular
way, which, however, also applies after relabeling $r=r(R)$ of the
level sets of the foliation of spacetime, yielding no more than the
trivial rescaling $N_{R}\rightarrow\frac{dr}{dR}N_{R}=:N_{r}$ of
the lapse. Accordingly, to include this very rescaling freedom, the
lapse function shall be chosen from now on as $N\equiv N_{r}$ for
determining the form of quasilocal quantities at the horizon. 

As a further step, let it be assumed that the horizon null vector
$l^{a}$ can be completed to a null tetrad of the form $(l^{a},k^{a},m^{a},\bar{m}^{a})$
such that the conditions $-l_{a}k^{a}=m_{a}\bar{m}^{a}=1$ are met.
Moreover, let it be assumed that there is a portion $\Delta\mathcal{T}$
of $\mathcal{T}$ that is bounded by two cross-sections $\mathcal{S}_{1}$
and $\mathcal{S}_{2}$, given with respect to the selected radial
coordinate $r(R)$, with radii $r_{1}=r(R_{1})$ and $r_{2}=r(R_{2})$
such that $r_{2}>r_{1}$. While the co-vector $k_{a}$ associated
with the transverse vector field $k^{a}$ is usually additionally
chosen as a null gradient field in the theory of dynamical and isolated
horizons, which has the consequence that $k^{a}$ is by definition
geodesic in these theories, this assumption is not needed and therefore
not made in the following. It could, however, still be made complementarily.

Anyway, with that set, it may be taken into account that the form
of the Brown-York Hamiltonian depends on which of the choices for
the time-flow vector field $t^{a}$ proposed in the previous section
is made in this context. Focusing here first on the case in which
$t^{a}=\sqrt{2}N_{r}l^{a}$, it is found that $H_{h}\equiv H_{h}^{Bulk}$
applies by necessity at $\mathcal{T}$ due to the fact that $\Theta=0$
holds along the same hypersurface. Consequently, the horizon part
of the Hamiltonian can be re-written in the form 
\begin{align}
H_{h} & =H_{h}^{Bulk}=\underset{\mathcal{T}}{\int}\mathscr{H}d^{3}x=\underset{\mathcal{T}}{\int}N_{r}(\mathcal{H}+\mathcal{H}_{a}s^{a})\omega_{h}=\\
 & =\sqrt{2}\underset{\mathcal{T}}{\int}N_{r}G_{ab}l^{a}n^{b}\omega_{h}=\underset{\mathcal{T}}{\int}N_{r}[G_{ab}l^{a}l^{b}+G_{ab}l^{a}k^{b}]\omega_{h}.\nonumber 
\end{align}
Hence, after using the decomposition $\mathcal{H}+\mathcal{H}_{a}s^{a}={}^{(2)}R-\sigma_{ab}\sigma^{ab}-2\zeta_{a}\zeta^{a}+\sqrt{2}\Theta(2K-\frac{3}{2}\Theta^{2})-\sqrt{2}\mathcal{L}_{s}\Theta$
with $\zeta^{a}:=q^{ab}(sD)l_{b}$ of the ADM Hamiltonian density,
this same Hamiltonian, as shown by Ashtekar and Krishnan in \cite{ashtekar2003dynamical},
gives rise to an energy flux term of the form
\begin{equation}
\mathcal{F}_{M}:=H_{h}\vert_{\Delta\mathcal{T}}=\frac{1}{16\pi}\underset{r_{1}}{\overset{r_{2}}{\int}}\underset{\mathcal{S}_{t}}{\int}({}^{(2)}R-\sigma_{ab}\sigma^{ab}-2\zeta_{a}\zeta^{a})\omega_{q}dr,
\end{equation}
where $^{(2)}R$ is the two-dimensional Ricci scalar. By taking the
Gauss-Bonnet theorem into account, it can then be shown that this
term leads for the standard choice $r(R)=R^{2}$ to an exact balance
law for the area increase of a given black hole, i.e.
\begin{equation}
\mathcal{A}_{2}-\mathcal{A}_{1}=\mathcal{F}_{M}+\mathcal{F}_{G},
\end{equation}
which is given with respect to the different black hole areas $\mathcal{A}_{j}=4\pi R_{j}^{2}$
with $j=1,2$, and an integral expression $\mathcal{F}_{G}=\underset{r_{1}}{\overset{r_{2}}{\int}}\underset{\mathcal{S}}{\int}_{t}(\sigma_{ab}\sigma^{ab}+2\zeta_{a}\zeta^{a})\omega_{q}dr$
describing the energy flux due to gravitational radiation. Based on
the fact that the right hand side of $(21)$ is manifestly non-negative,
this result thus shows that even in the fully dynamical case the area
of a black hole can never decrease.

Taking the above into account, it is found that the underlying ADM
Hamiltonian $H_{h}$ considered in $(19)$ can alternatively be cast
in the form 

\begin{equation}
H_{h}=\frac{1}{16\pi}\underset{\mathcal{T}}{\int}N_{r}(\frac{1}{2}{}^{(2)}R-\sigma_{ab}\sigma^{ab}-\Omega_{a}\Omega^{a})\omega_{h};
\end{equation}
thereby suggesting that $\underset{\mathcal{T}}{\int}N_{r}(\Omega_{a}\Omega^{a}+\frac{1}{2}{}^{(2)}R)\omega_{h}=2\underset{\mathcal{T}}{\int}N_{r}\zeta_{a}\zeta^{a}\omega_{h}$
is satisfied. This can be readily concluded from the fact that the
null Raychaudhuri equation $\mathfrak{L}_{l}\Theta=\kappa\Theta-\frac{1}{2}\Theta^{2}-\sigma_{ab}\sigma^{ab}+\omega_{ab}\omega^{ab}-G_{ab}l^{a}l^{b}$
can be combined with the identity $\mathfrak{L}_{k}\Theta=-\frac{1}{2}{}^{(2)}R-\Xi\Theta+\Omega_{a}\Omega^{a}-\mathcal{D}_{a}\Omega^{a}+G_{ab}l^{a}k^{b}$
to obtain the result
\begin{align}
\mathcal{H}+\mathcal{H}_{a}s^{a} & =\frac{1}{2}{}^{(2)}R+(\kappa+\Xi-\frac{1}{2}\Theta)\Theta-\sigma_{ab}\sigma^{ab}+\omega_{ab}\omega^{ab}-\\
 & -\Omega_{a}\Omega^{a}+\mathcal{D}_{a}\Omega^{a}-\sqrt{2}\mathcal{L}_{s}\Theta,\nonumber 
\end{align}
which can then be used to set up equation $(22)$. Note that it has
been used in this context that the expression $\int\left[\underset{\mathcal{S}_{t}}{\int}\mathcal{D}_{a}\Omega^{a}\omega_{q}\right]dr$
vanishes identically and $\underset{\mathcal{S}_{t}}{\int}\mathcal{D}_{a}\Omega^{a}\omega_{q}$
does so as well. As may be noted, the quantity $\kappa$ coincides
with the surface gravity of the black hole at the horizon, where one
generally has $\kappa=\epsilon+\bar{\epsilon}$ in spin-coefficient
notation.

With this clarified, it may next be noted that, although the boundary
term $H_{h}^{Boundary}$ is equal to zero, its temporal variation
with respect to the time evolution vector field $t^{a}$ is generally
not. Also, the corresponding variation of the bulk part $H_{h}^{Bulk}$
is generally unequal to zero, which has the consequence that the temporal
variation of the full Hamiltonian $H$ leads to an integral law with
a form identical to that of $(18)$ that includes a bulk-to-boundary
inflow term with integrand $(14)$ resulting from the variation of
the corresponding bulk part $H_{h}^{Bulk}$.

To see this, the Lie derivative of $H_{h}$ with respect to $t^{a}=\sqrt{2}N_{r}l^{a}$
along $\mathcal{T}$ will be calculated next. For this purpose, it
may be taken into account that a variation of the integrand occurring
in $(19)$ yields the result
\begin{align}
\mathfrak{L}_{t}[\omega_{h}N_{r}(\mathcal{H}+\mathcal{H}_{a}s^{a})] & =\sqrt{2}N_{r}\omega_{h}[\mathfrak{L}_{l}\ln N_{r}(\mathcal{H}+\mathcal{H}_{a}s^{a})+\mathfrak{L}_{l}(\mathcal{H}+\mathcal{H}_{a}s^{a})+\\
 & +(\Theta+\kappa)(\mathcal{H}+\mathcal{H}_{a}s^{a})]\nonumber 
\end{align}
where, just as a reminder, parts of the corresponding Hamiltonian
density can be written in the form $\mathcal{H}+\mathcal{H}_{a}s^{a}=G_{ab}l^{a}l^{b}+G_{ab}l^{a}k^{b}$.
Using here then the fact that

\begin{align}
\mathfrak{L}_{l}\left(G_{ab}l^{a}k^{b}\right) & =-\mathfrak{L}_{k}\left(G_{ab}l^{a}l^{b}\right)+G_{ab}(l\nabla)k^{a}l^{b}+2G_{ab}(k\nabla)l^{a}l^{b}+\\
 & +G_{ab}(l\nabla)l^{a}k^{b}+q_{a}^{\;c}\nabla_{c}G_{\;b}^{a}l^{b}\nonumber 
\end{align}
applies as a consequence of the contracted Bianchi identity $\nabla_{a}G_{\;b}^{a}\cdot l^{a}=0$,
it is thus found that 

\begin{align}
\mathfrak{L}_{l}[G_{ab}l^{a}l^{b}+ & G_{ab}l^{a}k^{b}]=\sqrt{2}\mathfrak{L}_{s}\left(G_{ab}l^{a}l^{b}\right)+\mathcal{D}_{a}(G_{\;b}^{a}l^{b})+G_{ab}\varkappa^{a}k^{b}+\\
+ & 2\tilde{\kappa}G_{ab}l^{a}l^{b}-2G_{ab}\tau^{a}l^{b}+2G_{ab}\Omega^{a}l^{b}+G_{ab}\sigma{}^{ab}+\frac{1}{2}\Theta G_{ab}q{}^{ab}\nonumber 
\end{align}
is satisfied, where, in spin-coefficient notation, one has $\tilde{\kappa}=\gamma+\bar{\gamma}$,
$\tau^{a}=\bar{\tau}m^{a}+\tau\bar{m}^{a}$, $\varkappa^{a}=\bar{\varkappa}m^{a}+\varkappa\bar{m}^{a}$
and $\Omega^{a}=(\bar{\alpha}+\beta)m^{a}+(\alpha+\bar{\beta})\bar{m}^{a}$.
Given that the co-vector $k_{a}$ is usually chosen as a gradient
field in the theory of dynamical and isolated horizons, one could
use at this point the fact that $\tilde{\kappa}=0$. Yet, since $(26)$
applies generically and remains valid regardless of whether $\tilde{\kappa}=0$
is satisfied or not, the latter is not strictly assumed either at
this or a later point of this work.

This being said, it can further be observed that
\begin{align}
N_{r}^{2}\mathfrak{L}_{s}\left(G_{ab}l^{a}l^{b}\right) & =\mathfrak{L}_{s}\left(N_{r}^{2}G_{ab}l^{a}l^{b}\right)-2N_{r}\mathfrak{L}_{s}N_{r}G_{ab}l^{a}l^{b}=D_{c}\left[(N_{r}^{2}G_{ab}l^{a}l^{b})s^{c}\right]-\nonumber \\
 & -[N_{r}^{2}\cdot k+2N_{r}\mathfrak{L}_{s}N_{r}]G_{ab}l^{a}l^{b}
\end{align}
applies globally in $\mathcal{M}$, where $k=\frac{1}{\sqrt{2}}(\Theta-\Xi)$
is the extrinsic curvature scalar calculated with respect to $s^{a}$.
Whence, using Gauss' law once again, one is thus led to conclude that
\begin{equation}
\underset{\mathcal{T}}{\int}D_{c}\left[(N_{r}^{2}G_{ab}l^{a}l^{b})s^{c}\right]\omega_{h}=\underset{\mathcal{S}_{t}}{\int}N_{r}^{2}G_{ab}l^{a}l^{b}\omega_{q}.
\end{equation}
This makes it clear that the bulk-to-boundary inflow term $(14)$
can be derived as required in the context of the theory of dynamical
black hole horizons. 

Clearly, since the Lie derivative of one and the same object is calculated
only in different ways, it can be confidently assumed that the remaining
terms in relations $(24-26)$ and $(28)$ can be combined to agree
with the integrand of the bulk integral term in equation $(6)$. However,
this confirms that the results of the quasilocal Brown-York-Hamiltonian
formalism and the dynamical horizon framework are entirely consistent
with each other; and that even to a much greater extent than pointed
out in \cite{booth2005horizon}.

For the sake of completeness, the result of the variation of the bulk
part of the Hamiltonian shall be given at this point as well, reading 

\begin{align}
\mathfrak{L}_{t}H_{h}^{Bulk} & =\frac{\sqrt{2}}{16\pi}\underset{\mathcal{T}}{\int}\omega_{h}N_{r}\{(\mathfrak{L}_{l}N_{r}+\kappa N_{r})({}^{(2)}R-\sigma_{ab}\sigma^{ab}-2\zeta_{a}\zeta^{a})+\\
 & +N_{r}(\mathfrak{L}_{l}{}^{(2)}R+\Omega_{ab}\sigma^{ab}-2\mathfrak{L}_{l}\vert\zeta\vert^{2})\},\nonumber 
\end{align}
where $\Omega_{ab}=q_{a}^{\;c}q_{b}^{\;d}C_{ecfd}l^{e}l^{f}$ and
$\vert\zeta\vert^{2}=\zeta_{a}\zeta^{a}$ applies by definition. Note
that the identity $\mathcal{L}_{l}\sigma_{ab}=q_{a}^{\;c}q_{b}^{\;d}L_{l}\sigma_{cd}=\kappa\sigma_{ab}+\sigma_{cd}\sigma^{cd}q_{ab}-q_{a}^{\;c}q_{b}^{\;d}C_{ecfd}l^{e}l^{f}$
has been used to obtain this form of relation $(29)$. An alternative
way to express this relation is

\begin{align}
\mathfrak{L}_{t}H_{h}^{Bulk} & =\frac{\sqrt{2}}{16\pi}\underset{\mathcal{T}}{\int}\omega_{h}N_{r}[(\mathfrak{L_{l}}N_{r}+\kappa N_{r})(G_{ab}l^{a}l^{b}+G_{ab}l^{a}k^{b})-N_{r}(\Xi-2\tilde{\kappa})G_{ab}l^{a}l^{b}+\nonumber \\
 & -N_{r}(N_{r}G_{ab}\varkappa^{a}k^{b}-2G_{ab}(\tau^{a}-\Omega^{a})l^{b}+G_{ab}\sigma{}^{ab}+2\mathfrak{L}_{l-k}N_{r}G_{ab}l^{a}l^{b})]+\nonumber \\
 & +\frac{1}{4\pi}\underset{\mathcal{S}_{t}}{\int}\omega_{q}N_{r}^{2}G_{ab}l^{a}l^{b},
\end{align}
thereby giving exactly the bulk-to-boundary inflow term derived in
the previous section. In this context, it has been used that

\begin{equation}
\underset{\mathcal{T}}{\int}N_{r}^{2}\mathcal{D}_{a}(G_{\;b}^{a}l^{b})\omega_{h}=\int drN_{r}\left[\underset{\mathcal{S}_{t}}{\int}\mathcal{D}_{a}(G_{\;b}^{a}l^{b})\omega_{q}\right]=0
\end{equation}
applies due to the fact that by Gauss' law the surface integral $\underset{\mathcal{S}}{\int}\mathcal{D}_{a}(G_{\;b}^{a}l^{b})\omega_{q}$
can be converted into an line integral over the boundary $\partial\mathcal{S}_{t}$
of $\mathcal{S}_{t}$ which is zero. 

However, as the variation of the Brown-York Hamiltonian $H_{h}$ with
respect to $t^{a}=\sqrt{2}N_{r}l^{a}$ at $\mathcal{T}$ is not yet
fully calculated, this is not the end of the story. To calculate the
latter, the variation $\mathcal{L}_{t}H_{h}^{Boundary}$ of the inner
boundary part of the Brown-York Hamiltonian has to be calculated as
well. Using once more the null Raychadhuri equation, one here finds 

\begin{align}
\mathcal{L}_{t}H_{h}^{Boundary} & =\\
= & \frac{1}{4\pi}\underset{\mathcal{S}_{t}}{\int}\left[N_{r}\mathcal{L}_{l}N_{r}\Theta+N_{r}^{2}(\kappa\Theta-\sigma_{ab}\sigma^{ab}+\omega_{ab}\omega^{ab}-8\pi T_{ab}l^{a}l^{b})\right]\omega_{q},\nonumber 
\end{align}
and thus
\begin{equation}
\mathcal{P}_{h}=\frac{1}{8\pi}\underset{\mathcal{S}_{t}}{\int}\mathcal{I}\omega_{q}=\frac{1}{4\pi}\underset{\mathcal{S}_{t}}{\int}\left[N_{r}\mathcal{L}_{l}N_{r}\Theta+N_{r}^{2}(\kappa\Theta-\sigma_{ab}\sigma^{ab}+\omega_{ab}\omega^{ab})\right]\omega_{q}
\end{equation}
for the power functional $\mathcal{P}_{h}$, as follows directly from
relations $(18)$ and $(30)$, respectively. Thus, as can be seen,
the bulk-to-boundary inflow term is counterbalanced by the net flow
of matter and/or radiation through the horizon. The amount of matter
and radiation that flows out through the horizon therefore flows back
in from the bulk, so that the net flux becomes zero at the cuts of
the horizon.

This result holds in the exact same form at $\Omega_{t}$, which is
interesting to the extent that, after taking the exterior boundary
of spacetime to future null infinity in the large sphere limit, it
is found that 

\begin{equation}
\mathcal{P}_{0}=-\frac{1}{4\pi}\underset{\mathbb{S}_{\infty}}{\int}N^{2}\sigma_{ab}\sigma^{ab}\omega_{q}=-\frac{1}{4\pi}\underset{\mathbb{S}_{\infty}}{\int}\vert n\vert^{2}d\Omega
\end{equation}
applies in a suitable Bondi-like chart $(u,r,\theta,\phi)$ with radial
null coordinate $r$ near null infinity \cite{bondi1962gravitational,Madler:2016xju}.
To see this, the asymptotic expansions $N[\mathcal{L}_{l}N+\kappa]\Theta\omega_{q}\underset{r\rightarrow\infty}{\longrightarrow}0$,
$N^{2}\omega_{ab}\omega^{ab}\omega_{q}\underset{r\rightarrow\infty}{\longrightarrow}0$
as well as $N^{2}\sigma_{ab}\sigma^{ab}\omega_{q}\underset{r\rightarrow\infty}{\longrightarrow}\vert n\vert^{2}d\Omega$
with $\sigma_{ab}\sigma^{ab}\omega_{q}=\frac{\vert n\vert^{2}}{r^{2}}r^{2}d\Omega$
may be taken into account, where $n(u,\theta,\phi)$ is the Bondi
news function. This news function is the retarded time derivative
of the radiation strain, i.e. $n=\partial_{u}\sigma_{0}$, where $\sigma_{0}$
corresponds to the leading order term of the spin-coefficient $\sigma$
of the Newman-Penrose formalism. This coefficient has been unobtrusively
incorporated into the definition of $(34)$ inasmuch as it has been
used that $\sigma_{ab}\sigma^{ab}=\vert\sigma\vert^{2}=\frac{\vert n\vert^{2}}{r^{2}}$,
where the latter holds in the vicinity of future null infinity. 

In light of the above, one is led to the conclusion that the quasilocal
quantity $\mathcal{P}_{0}$ given above coincides in Bondi coordinates
near future null infinity exactly with the time deriviative of the
Bondi mass aspect, thereby giving rise to the infamous Bondi mass-loss
formula \cite{bondi1962gravitational,Madler:2016xju}. This can readily
be concluded by taking into account that $\frac{dm_{B}}{du}=-\frac{1}{4\pi}\underset{\mathcal{S}_{\infty}}{\int}\vert n\vert^{2}d\Omega$
applies in said coordinates at future null infinity, where $m_{B}$
is the Bondi mass. However, this implies that the quantity $\mathcal{P}_{0}$
can be interpreted as one characterizing the rate of mass-loss of
a spatially and temporally bounded gravitating physical system, which
reduces to the standard expression given an extension of the spacetime
boundary to future null infinity. Yet, as may be noted, this only
holds true relative to the given choice $t^{a}\equiv\sqrt{2}Nl^{a}$
for the time evolution vector field, but not with respect to other
choices that do not lead to the same result.

To see this, let the general case $t^{a}=\sqrt{2}N_{r}l^{a}+V^{a}$
with $V^{a}=(O-N_{r})s^{a}+\mathcal{N}^{a}$ be considered, where
the situation changes completely in the sense that quasilocal corrections
to $\mathcal{P}_{0}$ and $\mathcal{P}_{h}$ occur naturally. Here,
the following is to be added: The choice $t^{a}=\sqrt{2}N_{r}l^{a}$
made above for the time flow vector field actually proves to be a
suitable choice for the study of the intrinsic and extrinsic geometry
of dynamical horizons and for characterizing the latter by means of
different null geometric quantities. The choice of an alternative
time-flow vector field is essentially arbitrary for spacetimes lacking
time translation symmetry; however, this choice should always be made
taking into account important local geometric properties of the considered
geometric field. On the other hand, given spacetimes that do not lack
time translation (and/or rotational) symmetry, it is straightforward
to make a choice for the time flow vector field of the geometry. This
choice, as already emphasized in the previous section, is simply given
by the Killing vector field of spacetime , i.e. by the linear combination
of temporal and angular Killing vector fields, which can be written
the form $t^{a}=\sqrt{2}Nl^{a}+V^{a}$ with $V^{a}=\Omega\varphi^{a}$.
Such a choice already leads to quasilocal corrections, as shall be
shown below.

To derive said corrections and obtain a bulk-to-boundary inflow term
of the form $(12)$, one may perform a variation of the bulk part
of the horizon Hamiltonian. From the perspective of an unboosted observer
(for which $\eta=0$), this Hamiltonian takes the form

\begin{equation}
H_{h}=H_{h}^{Bulk}+H_{h}^{Boundary}=\underset{\mathcal{T}}{\int}(\mathscr{H}+\mathscr{R})d^{3}x+\underset{\mathcal{S}_{t}}{\int}\mathcal{\mathfrak{H}}d^{2}x
\end{equation}
where the definitions $\mathscr{H}:=\frac{\sqrt{h}}{8\pi}N_{r}(\mathcal{H}+\mathcal{H}_{a}s^{a})$
and $\mathscr{R}:=\frac{\sqrt{h}}{8\pi}\mathcal{H}_{a}V^{a}=\frac{\sqrt{h}}{8\pi}[(O-N_{r})\mathcal{H}_{a}s^{a}+\mathcal{H}_{a}\mathcal{N}^{a}]$
have been used. Perhaps the simplest way to calculate the variation
$\mathfrak{L}_{t}H_{h}^{Bulk}$ of the bulk part is to consider the
$3+1$-identities

\begin{align}
\mathfrak{L}_{V}H_{h}^{Bulk} & =\frac{1}{8\pi}\underset{\mathcal{T}}{\int}D_{b}\{[N_{r}(\mathcal{H}+\mathcal{H}_{a}s^{a})+\mathcal{H}_{a}V^{a}]V^{b}\}\omega_{h}=\\
 & =\frac{1}{8\pi}\underset{\mathcal{S}_{t}}{\int}[N_{r}(O-N_{r})(\mathcal{H}+\mathcal{H}_{a}s^{a})+\mathcal{H}_{a}V^{a}]\omega_{q}\nonumber 
\end{align}
and

\begin{align}
\nabla_{c}G_{\;b}^{c}V^{b} & =-\mathfrak{L}_{n}\mathcal{H}_{b}V^{b}+\mathcal{H}a_{b}V^{b}-K\mathcal{H}_{b}V^{b}+\\
 & +\mathcal{Q}_{\;b}^{c}a_{c}V^{b}+D_{c}\mathcal{Q}_{\;b}^{c}V^{b}=0,\nonumber 
\end{align}
where the fact that $V_{a}s^{a}=O-N_{r}$ has been used. As may be
noted, the latter identity can straightforwardly be deduced using
the $3+1$-splitting $G_{\;b}^{a}=\mathcal{H}n^{a}n_{b}-\mathcal{H}^{a}n_{b}-n^{a}\mathcal{H}_{b}+\mathcal{Q}_{\;b}^{a}$
of the Einstein tensor. 

Taking into account the decomposition relations $n^{a}=\frac{1}{\sqrt{2}}(l^{a}+k^{a})$,
relation $(37)$ can be recast such that

\begin{equation}
\mathcal{L}_{l}\mathcal{H}_{b}V^{b}=-\mathcal{L}_{k}\mathcal{H}_{b}V^{b}+D_{c}(\mathcal{Q}_{\;b}^{c}V^{b})+\Phi_{V}
\end{equation}
with

\begin{equation}
\Phi_{V}=\mathcal{H}V^{b}D_{b}\ln N_{r}-K\mathcal{H}_{b}V^{b}+\mathcal{Q}_{\;b}^{c}V^{b}D_{c}\ln N_{r}-\mathcal{Q}_{\;b}^{c}D_{c}V^{b}
\end{equation}
being satisfied. Ultimately, applying Gauss' theorem once again to
convert
\begin{equation}
\underset{\mathcal{T}}{\int}D_{c}(N_{r}\mathcal{Q}_{\;b}^{c}V^{b})\omega_{h}=\underset{\mathcal{S}_{t}}{\int}\mathcal{Q}_{\;b}^{c}s_{c}V^{b}\omega_{q}
\end{equation}
and taking $(30)$ as well as $(36)$ and $(39)$ into account, one
finds that for a time-flow vector field of the form $t^{a}=\sqrt{2}N_{r}l^{a}+V^{a}$
the variation $\mathfrak{L}_{t}H_{h}^{Bulk}$ of the bulk part of
the Hamiltonian is given by $(30)$ plus an extra term of the form
\begin{align}
\frac{1}{8\pi}\underset{\mathcal{T}}{\int}\sqrt{2} & N_{r}[\mathcal{L}_{l}\mathcal{H}_{b}V^{b}+\mathcal{H}_{b}\mathcal{L}_{l}V^{b}+]\omega_{h}+\mathfrak{L}_{V}H_{h}^{Bulk}=\\
=\frac{1}{8\pi} & \underset{\mathcal{T}}{\int}[-\mathcal{L}_{k}\mathcal{H}_{b}V^{b}+\mathcal{H}_{b}\mathcal{L}_{l}V^{b}-\mathcal{Q}_{\;b}^{c}V^{b}D_{c}N_{r}+\Phi_{V}]\omega_{h}+\underset{\mathcal{S}_{t}}{\int}\varPi_{V}\omega_{q}\nonumber 
\end{align}
with 
\begin{equation}
\varPi_{V}=\frac{1}{8\pi}\{(O-N_{r})[N_{r}(\mathcal{H}+\mathcal{H}_{a}s^{a})+\mathcal{H}_{a}V^{a}]+N_{r}\mathcal{Q}_{\;b}^{c}s_{c}V^{b}\}.
\end{equation}
As can readily be checked, by combining $(30)$ and $(41)$ one obtains
again a bulk-to-boundary inflow term with an integrand that satisfies
the identity $\varPi+\varPi_{V}=\Pi=\Pi_{0}+\varPi_{N}$. Thus, as
to be expected, one finds the result of the first section exactly
reproduced.

For the variation of the boundary part of the Hamiltonian $H_{h}^{Boundary}$,
one may take into account that $\mathcal{L}_{\mathcal{N}}\underset{\mathcal{S}}{\int}(\sqrt{2}N_{r}\Theta-\varGamma_{V})\omega_{q}=\underset{\mathcal{S}}{\int}\mathcal{D}_{c}[(\sqrt{2}N_{r}\Theta-\varGamma_{V})\mathcal{N}^{c}]\omega_{q}=0$
applies in the given context. This reveals the fact that the variation
$\mathcal{L}_{t}H_{h}^{Boundary}$ of the boundary term is given by
$(32)$ plus an extra term of the form
\begin{equation}
\underset{\mathcal{S}_{t}}{\int}\Delta_{V}\omega_{q}=\frac{1}{8\pi}\underset{\mathcal{S}_{t}}{\int}\sqrt{2}N_{r}[\mathcal{L}_{l}\varGamma_{V}+\Theta\varGamma_{V}-\Psi_{V}]
\end{equation}
where the definition

\begin{align}
\Psi_{V} & =(O-N_{r})[\sqrt{2}\mathcal{L}_{s}N_{r}\Theta+\sqrt{2}N_{r}\mathcal{L}_{s}\Theta+(\sqrt{2}N_{r}\Theta-\varGamma_{V})k-\mathcal{L}_{s}\varGamma_{V}]=\\
=N_{r} & (O-N_{r})[\mathcal{L}_{l-k}\ln N_{r}\cdot\Theta+\frac{1}{2}{}^{(2)}R+(\kappa+\Xi-\frac{1}{2}\Theta)\Theta+\frac{1}{\sqrt{2}}(\sqrt{2}N_{r}\Theta-\varGamma_{V})(\Theta+\Xi)-\nonumber \\
 & -\sigma_{ab}\sigma^{ab}+\omega_{ab}\omega^{ab}-\Omega_{a}\Omega^{a}+\mathcal{D}_{a}\Omega^{a}-\mathcal{H}-\mathcal{H}_{a}s^{a}-\frac{1}{\sqrt{2}}\mathcal{L}_{l-k}\varGamma_{V}]\nonumber 
\end{align}
has been used, which takes a simpler form at $\mathcal{T}$ and therefore
also at $\mathcal{S}_{t}$ due to the fact that $\Theta=0$ applies
there. Hence, by combining the bulk-to-boundary terms resulting from
$(30)$ and $(41)$ with $(43)$, one obtains the power functional
\begin{equation}
\mathscr{P}_{h}=\mathcal{P}_{h}+\mathfrak{P}_{h},
\end{equation}
for the inner boundary of spacetime, with $\mathcal{P}_{h}$ being
depicted in $(33)$. This power functional contains the quasilocal
correction term 
\begin{equation}
\mathfrak{P}_{h}=\underset{\mathcal{S}_{t}}{\int}(\varPi_{V}-\Delta_{V})\omega_{q},
\end{equation}
 which again takes a simpler form at $\mathcal{S}_{t}$ due to the
fact that $\Theta=0$ is satisfied along the leaves of $\mathcal{T}.$
The reason why $\Theta$ was not set equal to zero in this context
is, of course, that by an analogous approach an equivalent power functional

\begin{equation}
\mathscr{P}_{0}=\mathcal{P}_{0}+\mathfrak{P}_{0},
\end{equation}
can be derived for the outer boundary of spacetime, whose from can
easily be read off from $(45)$. In fact, the exact same expression
is obtained here, only that $\mathcal{S}_{t}$ has to be replaced
by $\Omega_{t}$ and it cannot necessarily be assumed that $\Theta$
is equal to zero, since the outer horizon is a timelike hypersurface
that will not generally represent a dynamical horizon. Provided that,
as before, the outer boundary of spacetime is shifted to null infinity
in the large sphere limit, $\mathcal{P}_{0}$ is then again given
by equation $(34)$, thereby implying that $\mathfrak{P}_{0}$ describes
quasilocal corrections to said formula. Note that here again the transition
to the rotating case can be readily achieved by the substitutions
$N\rightarrow\tilde{N}$, $O\rightarrow\tilde{O}$, and $\mathcal{N}^{a}\rightarrow\tilde{\mathcal{N}}^{a}$.

As can be inferred from equations $(45)$ and $(47)$, the quantities
$\mathfrak{P}_{h}$ and $\mathfrak{P}_{0}$ encode corrections to
the quasilocal analog of Bondi's mass-loss formula given by equations
$(33)$ and $(34)$, respectively. These corrections each contain
one term resulting from the variation of the boundary part of the
Hamiltonian and a previously unrecognized bulk-to-boundary inflow
term. The existence of precisely these terms suggests that the Bondi
mass-loss formula - for the given choice of the time evolution vector
field - results as a special case of the Brown-York formalism only
when $\mathfrak{P}_{0}\underset{r\rightarrow\infty}{\longrightarrow}0$
is satisfied, which, however, requires specific asymptotic fall-off
conditions to be satisfied. Yet, the latter may not be the case for
all choices $N^{a}=Os^{a}+\mathcal{N}^{a}$ for the shift vector of
the geometry; possibly not even in case that $N^{a}=0$, as will be
explained more clearly in the following, final section of this paper.

In conclusion, given the fact that in general no preferred time-flow
vector field can be distinguished in spacetimes which do not exhibit
time-translation symmetry and, moreover, with $O$ and $\mathcal{N}^{a}$
essentially freely selectable quantities which enter the definition
of the derived qusilocal corrections, it is clear that these quantities
can always be chosen in such a way that the mentioned corrections
are non-zero in spacetimes without the mentioned symmetry. Thus, one
is led to conclude that, according to the quasilocal Hamiltonian formalism
used in this work, there are integral terms that can be expected to
survive the large sphere limit and so give rise to non-vanishing quasilocal
corrections to Bondi's mass-loss formula. Accordingly, there should
be exceptions to the generally accepted rule: \textit{The mass of
a system is constant if and only if there is no news}. Rather, in
light of the above, the statement should read correctly: \textit{The
mass of a system is constant if and only if there is no news (and
the bulk stress-energy tensor at the boundary of spacetime is zero)}.

This small addition to Bondi's statement proves quite significant
in some cases of interest, in particular those in which both electromagnetic
and gravitational radiation escape continuously from the system to
null infinity, which is possible in non-stationary spacetimes because
both matter and radiation can simultaneously pass through a dynamical
horizon (as opposed to the case of a stationary isolated or even Killing
horizon, where the latter would be impossible). The reason for this
is that in just such a case there should be a non-vanishing bulk-to-boundary
energy inflow term and hence corrections to the Bondi mass-loss formula,
which should manifest themselves in the form of a shift in the intensity
of the measured radiation. 

That the explicit form of such a shift can actually be calculated,
at least in simpler cases, will be shown in the forthcoming concluding
section of this work, using the Generalized Vaidya family of spacetimes
as an example. In doing so, it will be shown that the quasilocal corrections
derived in the present section must be taken into account in order
to obtain the correct formula for the loss of mass and radiant energy
through a dynamical horizon in a spatially and temporally bounded
gravitational system whose boundary is shifted to infinity in the
large sphere limit. The form of the resulting quasilocal corrections
is shown to depend particularly on the choice of the shift vector
field of the geometry; with certain choices clearly showing that,
contrary to general expectation, null radiation at null infinity can
be detected even when both the Bondi news function and the time derivative
of the associated mass aspect are zero. 

\section{A Specific Example: The Family of Generalized Vaidya Spacetimes}

To illustrate applications for the integral laws derived in the previous
sections, a specific class of geometric models will now be treated
in the following, namely the Generalized Vaidya family of solutions
of Einstein's eqations. The metric of this spacetime describes the
geometric field of a matter distribution with a null dust and a non-rotating
null fluid part. In the ingoing case, the line element encoding the
components corresponding metric reads
\begin{equation}
ds^{2}=-(1-\frac{2M}{r})dv^{2}+2dvdr+r^{2}(d\theta^{2}+\sin^{2}\theta d\phi^{2}),
\end{equation}
where $M(v,r)$ is assumed to be a function of $v$ and $r$ with
the property that its first derivative with respect to $v$ possesses
a well-defined limit in the sense that the object $\dot{M}_{\infty}=\underset{v,r\rightarrow\infty}{\lim}\dot{M}$
with $\dot{M}=\partial_{v}M$ exists and is non-singular. Given that
this is the case, one may consider the normalized geodesic null frame

\begin{align}
l^{a} & =\partial_{v}^{a}+\frac{N^{2}}{2}\partial_{r}^{a}\\
k^{a} & =-\partial_{r}^{a}\nonumber \\
m^{a} & =\frac{1}{\sqrt{2}r}(\partial_{\theta}^{a}+i\csc\theta\partial_{\phi}^{a})\nonumber \\
\bar{m}^{a} & =\frac{1}{\sqrt{2}r}(\partial_{\theta}^{a}-i\csc\theta\partial_{\phi}^{a}),\nonumber 
\end{align}
in relation to which the stress-energy tensor of this geometry splits
in two parts
\begin{equation}
T_{ab}=T_{ab}^{(D)}+T_{ab}^{(F)},
\end{equation}
i.e., a null dust part $T_{ab}^{(D)}=\mu k_{a}k_{b}$ of type $I$
and a null fluid part $T_{ab}^{(F)}=2(\rho+p)l_{(a}k_{b)}+pg_{ab}$
of type $II$ \cite{hawking2023large}, where the short hand notation
$\mu=\frac{\dot{M}}{4\pi r^{2}}$, $\rho=\frac{M'}{4\pi r^{2}}$,
$p=-\frac{M''}{8\pi r}$ , $M':=\partial_{r}M$, $k_{a}=dv_{a}$ and
$l_{a}=dr_{a}-\frac{1}{2}(1-\frac{2M}{r})dv_{a}$ has been introduced.
Thus, given that $M$ is only a function of $v$, $T_{ab}^{(F)}$
is zero and the Vaidya metric is obtained as a special case. On the
other hand, if $M$ is chosen to be of the form $M(v,r)=m(v)+\frac{\Lambda r^{3}}{6}$
the Vaidya-de Sitter metric is obtained as a special case, which reduces
to the Kottler alias Schwarzschild-de Sitter metric in the case that
$m(v)=m_{0}=const.$

As a basis for the introduction of a geometric setting, as considered
in the first section, one may now identify $N^{2}=1-\frac{2M}{r}$
as the lapse function of the geometry and perform a rescaling of the
form $l^{a}\rightarrow\frac{\sqrt{2}}{N}l^{a}$ and $k^{a}\rightarrow\frac{N}{\sqrt{2}}k^{a}$,
which yields the timelike and spacelike vector fields $n^{a}=\frac{1}{\sqrt{2}N}\partial_{v}^{a}$
and $s^{a}=\frac{1}{\sqrt{2}}[\frac{1}{N}\partial_{v}^{a}+N\partial_{r}^{a}]$.
Then, by another boost transformation, the related vector fields $v^{a}=\frac{1}{\lambda}n^{a}-\eta s^{a}$
and $u^{a}=\frac{1}{\lambda}s^{a}-\eta n^{a}$ can be constructed
in the next step, so that the main ingredients for the introduction
of the geometric setting considered in previous sections of this work
are given. Taking then further into account that the generalized Vaidya
geometry is non-stationary and thus lacks time translation symmetry,
it becomes clear that there are various ways to select the time-flow
vector field of spacetime, all of which are consistent with the results
of the previous section. As a result, there are multiple ways to set
up the part $H_{0}$ of the quasilocal Hamiltonian $H$, and also
multiple ways to calculate the variation of this same Hamiltonian
and to investigate whether quasilocal corrections arise and persist
even when the boundary of spacetime is shifted to infinity; some of
which will now be discussed in the following.

Obviously, one of the choices mentioned is $t^{a}=\sqrt{2}Nl^{a}$.
Since for this choice $\kappa=-\mathcal{L}_{l}N$ and $\omega_{ab}=0$
applies, it is clear that the asymptotic fall-off conditions $N[\mathcal{L}_{l}N+\kappa]\Theta\omega_{q}\underset{r\rightarrow\infty}{\longrightarrow}0$,
$N^{2}\omega_{ab}\omega^{ab}\omega_{q}\underset{r\rightarrow\infty}{\longrightarrow}0$
are met in the case that the boundary of spacetime is moved to null
infinity in the large sphere limit. 

Thus, taking the results of the previous section into account, it
follows that the associated power functional $\mathcal{P}_{0}$, which
describes the rate at which energy is radiated to infinity, is given
by relation $(34)$ and that only the energy flux due to gravitational
radiation reaches null infinity; thereby proving to be consistent
with relation $(47)$ in the sense that $\mathfrak{P}_{0}=0$. Hence,
no quasilocal corrections arise in the given case. Yet, since the
generalized Vaidya metric is spherically symmetric, it is clear that
$\sigma_{ab}=0$ and thus $\mathcal{P}_{0}=\frac{dm}{dv}=0$ applies,
thereby implying that the Bondi mass $m$ of the system is constant,
as to be expected.

Another possible choice for the time-flow vector field of spacetime
is $t^{a}=\frac{1}{\sqrt{2}}\partial_{v}^{a}=Nn^{a}=\frac{N}{\sqrt{2}}(l^{a}+k^{a})$;
a choice according to which $N^{a}=0$ applies. Given this comparatively
simple form of $t^{a}$, it turns out to be most straightforward to
use $(10)$ directly to determine the form of potentially occurring
quasilocal corrections. To this end, it may first be concluded from
$(9)$ that the bulk-to-boundary inflow term of the geometry takes
the form 
\begin{align}
\underset{\Omega_{v}}{\int}\Pi\omega_{q} & =\frac{1}{8\pi}\underset{\Omega_{v}}{\int}N^{2}G_{ab}n^{a}s^{b}\omega_{q}=\frac{1}{2}\underset{\Omega_{v}}{\int}N^{2}T_{ab}l^{a}l^{b}\omega_{q}=\frac{1}{2}\underset{\Omega_{v}}{\int}N^{2}\mu\omega_{q}=\\
 & =\frac{1}{2}(1-\frac{2M}{r})\dot{M},\nonumber 
\end{align}
thereby yielding 
\begin{equation}
\underset{\mathbb{S}_{\infty}}{\int}\Pi\omega_{q}=\frac{1}{2}\dot{M}_{\infty}
\end{equation}
in the large sphere limit. Then, given that the validity of $N^{a}=0$
implies that $O=\mathcal{N}^{a}=0$ and thus $\varGamma_{V}=\sqrt{2}N\Theta-Nk$
is satisfied, one finds that

\begin{equation}
\mathcal{\mathfrak{H}}=\frac{\sqrt{q}}{8\pi}Nk.
\end{equation}
Also, by taking further into account that $k=\frac{1}{\sqrt{2}}(\Theta-\Xi)$,
$\Theta=\frac{N^{2}}{r}$ and $\Xi=-\frac{2}{r}$ and thus $\Theta-\Xi=\frac{N^{2}+2}{r}$
applies in the given context, it is found that
\begin{align}
\frac{dH_{0}^{Boundary}}{dt} & =\frac{1}{\sqrt{2}8\pi}\underset{\Omega_{v}}{\int}N\mathcal{L}_{n}N(\Theta-\Xi)+N^{2}\mathcal{L}_{n}(\Theta-\Xi)+\frac{N^{2}}{\sqrt{2}}(\Theta^{2}-\Xi^{2})]\omega_{q}=\\
 & =\frac{1}{16\pi}\underset{\Omega_{v}}{\int}[\frac{(3N^{2}+2)\partial_{v}N^{2}}{2Nr}+\frac{N^{2}(N^{4}-4)}{r^{2}}]\omega_{q},\nonumber 
\end{align}
where, just as a reminder, $N^{2}=1-\frac{2M}{r}$ applies by definition.
Ultimately, by using the fact that$\frac{1}{2}\partial_{v}N^{2}=-\frac{\dot{M}}{r}$
and $\omega_{q}=r^{2}\sin\theta d\theta d\phi$ holds true by definition
and taking the large sphere limit of $(54)$, one obtains the final
result
\begin{equation}
\mathscr{P}_{0}=\mathfrak{P}_{0}=\underset{v,r\rightarrow\infty}{\lim}\frac{dH_{0}^{Boundary}}{dt}+\underset{\mathbb{S}_{\infty}}{\int}\Pi\omega_{q}=\frac{3}{4}(1-\dot{M}_{\infty})\neq0.
\end{equation}
From this, however, it can be concluded that even in the given simple
case Bondi's result is no longer exactly reproduced, since the variation
of the quasilocal mass $m$ of the system does not coincide with that
of the Bondi mass $m_{B}$; the fact that both quasilocal masses coincide
exactly at future null infinity notwithstanding.

Moreover, since neither $T_{ab}n^{a}n^{b}$ nor $T_{ab}n^{a}s^{b}$
nor $T_{ab}s^{a}s^{b}$ are equal to zero, it is clear that one can
always find a function $O(v,r)$ and an associated time evolution
vector field of the form $t^{a}=\sqrt{2}Nl^{a}+(O-N)s^{a}+\mathcal{N}^{a}$
such that the resulting bulk-to-boundary inflow term with integrand
$(12)$ is different from zero at null infinity, thereby implying
that $\mathfrak{P}_{0}\neq0$ and thus $\mathscr{P}_{0}\neq0$ applies
for the quasilocal quantities occurring in $(47)$ in such a case.
Accordingly, given this particular choice of time-flow vector field,
it becomes clear that, as claimed, quasilocal corrections to the Bondi
mass-loss formula occur at future null infinity. These are manifestly
different from zero, so that in such a case one is led to conclude
that there are again deviations from Bondi's mass-loss formula caused
by the resulting quasilocal corrections. 

Thus, to conclude, it is found in the given non-stationary case that
radiation fields can be detected at null infinity even in cases where
the Bondi news function is zero. The bulk-to-boundary inflow term
responsible for this fact depends to a large extent, similar to the
other quasilocal corrections, on the choice of the time-flow vector
field of the geometry. For special choices of the latter, as it turns
out, Bondi's results can however be exactly reproduced.

That said, in order to determine the analogous power functional at
the horizon, one may proceed somewhat differently; if only because
the rescaled null vector field $l^{a}=\frac{\sqrt{2}}{N}\partial_{v}^{a}+\frac{N}{\sqrt{2}}\partial_{r}^{a}$
diverges at the horizon.

A facilitating circumstance in this context is the fact that it proves
sufficient to consider vector fields which are only locally lightlike
at the dynamical horizon $\mathcal{T}$. Taking this into account,
the steps taken in \cite{ashtekar2003dynamical} can be applied as
is to the given case and used as a basis for constructing the horizon
Hamiltonian $H_{h}$ as well as calculating its variation. To this
end, one may define the function $f(v,r):=N^{2}(v,r)=1-\frac{2M(v,r)}{r}$,
choose $N_{r}=\sqrt{\vert\frac{\dot{f}}{2f'}\vert}=\sqrt{\frac{1}{2}\frac{\dot{M}r}{M-rM'}}$
for the lapse function of the geometry and set up the system of vector
fields $\tilde{l}^{a}=\frac{\vert f'\vert}{\sqrt{\vert\dot{f}f'\vert}}\partial_{v}^{a}$,
$\tilde{k}^{a}=-\frac{\vert\dot{f}\vert}{\sqrt{\vert\dot{f}f'\vert}}\partial_{r}^{a}$
$\tilde{n}^{a}=\frac{1}{\sqrt{\vert2\dot{f}f'\vert}}[f'\partial_{v}^{a}+\dot{f}\partial_{r}^{a}]$,
$\tilde{s}^{a}=\frac{1}{\sqrt{\vert2\dot{f}f'\vert}}[f'\partial_{v}^{a}-\dot{f}\partial_{r}^{a}]$,
with $\tilde{l}^{a}$ being only locally lightlike. Choosing then
the horizon vector field $\tilde{t}^{a}=\sqrt{2}N_{r}\tilde{l}^{a}$
for setting up $H_{h}$\footnote{In this context, it has to be ensured that the horizon vector field
transitions smoothly into the residual time-flow vector field. However,
this can readily be done by considering a smooth joining of the two
vector fields using distrbutions or functions with compact support.}, it can be concluded from $(33)$ that $\mathscr{P}_{h}=\mathcal{P}_{h}=\mathfrak{P}_{h}=0$,
which implies that the gravitational Hamiltonian of generalized Vaidya
spacetime, when defined with respect to $\tilde{t}^{a}$, is constant
with respect to the Lie-flow generated by the same vector field, and
thus is a locally conserved quantity at the boundary of spacetime
(but not in the bulk). In the case where $M\rightarrow const.$ and
thus the geometry of spacetime approaches that of Schwarzschild spacetime,
the fact that $N_{r}\rightarrow0$ applies in the same limit further
implies that $H_{h}\rightarrow0$, as to be expected. 

Consequently, given the choice $N_{r}$ for the lapse function at
the horizon, the fact that $H_{h}\rightarrow0$ shows that the temporal
variation of the total quasilocal Brown-York Hamiltonian vanishes
once the black hole horizon reaches a steady state of equilibrium
and becomes an isolated or weakly isolated horizon in the sense of
Ashtekar et al. Thus, the treated model confirms that any matter and/or
radiation flux (of the specified type) from the bulk to the boundary
of spacetime crossing a dynamical horizon necessarily subsides completely
in the limiting case where the geometry of the generalized Vaidya
becomes static and settles into an equilibrium state where it coincides
with that of Schwarzschild spacetime. The model therefore confirms
that in the case of a black hole spacetime, neither matter nor radiation
can escape to infinity through the event horizon of a black hole.
However, through a dynamical horizon, which lies not within a black
hole event horizon, matter and radiation can very well escape to infinity.
An example of the occurrence of such a situation is in the case of
Vaidya-de Sitter spacetime for $M>\sqrt{9\Lambda}$; a case in which
no black hole horizon can form, but spacetime nevertheless exhibits
a dynamical horizon through which matter and radiation can escape
to infinity (but not to future null infinity, as the latter does not
exist in said case). But there are, of course, other examples that
could also be mentioned at this point.

Anyway, the above should apply not only to dynamical horizons whose
cross sections are spherically symmetric, but also to more general
horizons that occur, for example, in non-stationary axisymmetric spacetime.
However, surprisingly, it has been found in the literature that in
general it is not easy to analyze non-spherical dynamical horizons,
since not too much is known about non-spherical marginally trapped
surfaces. 

Yet, as far as the results of the present work are concerned, this
does not pose a major problem, since the derived integral laws should
retain their validity for any type of dynamical black hole spacetime
(and certainly beyond); even if the notion of a dynamical horizon
is replaced by Hayward's more general notion of a trapping horizon.
The calculated quasilocal corrections should therefore be taken into
account where necessary.

\section*{Conclusion and Outlook}

In this work, the rate of change of mass and/or radiant energy escaping
through the spatial boundary of a confined non-stationary spacetime
was calculated using the quasilocal Brown-York formalism. In doing
so, it was shown that a null geometric equivalent of the bulk-to-boundary
inflow term derived in \cite{huber2022remark} results from varying
the total Hamiltonian of the theory, which describes how matter and/or
radiation can escape from the bulk of spacetime into its boundary
region. Also, it was shown that other quasilocal corrections occur,
some of which do not vanish even when the boundary of spacetime is
shifted to infinity. As a result, using the example of Generalized
Vaidya spacetime, it was shown that, in general, corrections to the
Bondi mass-loss formula occur at null infinity, even though said formula
can also be reproduced exactly - given a suitable choice of the time-flow
vector field. The null geometric approach used for this purpose was
found to be consistent with the theory of dynamical and isolated horizons,
and it was found that the horizon part of the Hamiltonian becomes
zero (for a suitable choice for the lapse function of the geometry)
as soon as the dynamical horizon transitions into an isolated or weakly
isolated horizon. 

Remarkably, in this context, it turns out that the form of the derived
bulk-to-boundary inflow term is independent of the choice of boundary
conditions chosen to set up the quasilocal Hamiltonian of the theory.
The reason is that this term results from the variation of the bulk
part of the quasilocal Hamiltonian, i.e., the ADM Hamiltonian, and
not from the variation of its boundary part. For this reason, the
quasilocal corrections associated with this term always occur when
the ADM Hamiltonian is considered, and are thus relatively universally
applicable in general relativity. It is therefore to be expected that
said quasilocal corrections play an important role in describing a
large class of phenomena in Einstein-Hilbert gravity. As it stands,
any further applications will be discussed in more detail elsewhere,
in a future work on this subject.
\begin{description}
\item [{Acknowledgements:}]~
\end{description}
Great thanks to Abhay Ashtekar for pointing out an erroneous conclusion
in the first draft of the manuscript. Also, I want to thank Felix
Wilkens for his support in preparing the image depicted in Fig. 1
of the paper.

\bibliographystyle{plain}
\addcontentsline{toc}{section}{\refname}\bibliography{1C__Arbeiten_litquas2}

\end{document}